\documentstyle[psfig]{mn}

\newif\ifAMStwofonts

\ifoldfss
  \ifCUPmtlplainloaded \else
    \NewTextAlphabet{textbfit} {cmbxti10} {}
    \NewTextAlphabet{textbfss} {cmssbx10} {}
    \NewMathAlphabet{mathbfit} {cmbxti10} {} 
    \NewMathAlphabet{mathbfss} {cmssbx10} {} 
  \fi
  \ifAMStwofonts
    \ifCUPmtlplainloaded \else
      \NewSymbolFont{upmath} {eurm10}
      \NewSymbolFont{AMSa} {msam10}
      \NewMathSymbol{\upi}     {0}{upmath}{19}
      \NewMathSymbol{\umu}     {0}{upmath}{16}
      \NewMathSymbol{\upartial}{0}{upmath}{40}
      \NewMathSymbol{\leqslant}{3}{AMSa}{36}
      \NewMathSymbol{\geqslant}{3}{AMSa}{3E}

    \fi
  \fi
\fi 

\ifnfssone
  \newmathalphabet{\mathit}
  \addtoversion{normal}{\mathit}{cmr}{m}{it}
  \addtoversion{bold}{\mathit}{cmr}{bx}{it}
  \newmathalphabet{\mathbfit} 
  \addtoversion{normal}{\mathbfit}{cmr}{bx}{it}
  \addtoversion{bold}{\mathbfit}{cmr}{bx}{it}
  \newmathalphabet{\mathbfss} 
  \addtoversion{normal}{\mathbfss}{cmss}{bx}{n}
  \addtoversion{bold}{\mathbfss}{cmss}{bx}{n}
  \ifAMStwofonts
    \ifCUPmtlplainloaded \else
      \UseAMStwoboldmath
      \makeatletter
      \new@mathgroup\upmath@group
      \define@mathgroup\mv@normal\upmath@group{eur}{m}{n}
      \define@mathgroup\mv@bold\upmath@group{eur}{b}{n}
      \edef\UPM{\hexnumber\upmath@group}
      \new@mathgroup\amsa@group
      \define@mathgroup\mv@normal\amsa@group{msa}{m}{n}
      \define@mathgroup\mv@bold\amsa@group{msa}{m}{n}
      \edef\AMSa{\hexnumber\amsa@group}
      \makeatother
      \mathchardef\upi="0\UPM19
      \mathchardef\umu="0\UPM16
      \mathchardef\upartial="0\UPM40
      \mathchardef\leqslant="3\AMSa36
      \mathchardef\geqslant="3\AMSa3E
    \fi
  \fi
\fi 

\ifnfsstwo
  \DeclareMathAlphabet{\mathbfit}{OT1}{cmr}{bx}{it}
  \SetMathAlphabet\mathbfit{bold}{OT1}{cmr}{bx}{it}
  \DeclareMathAlphabet{\mathbfss}{OT1}{cmss}{bx}{n}
  \SetMathAlphabet\mathbfss{bold}{OT1}{cmss}{bx}{n}
  \ifAMStwofonts
    \ifCUPmtlplainloaded \else
      \DeclareSymbolFont{UPM}{U}{eur}{m}{n}
      \SetSymbolFont{UPM}{bold}{U}{eur}{b}{n}
      \DeclareSymbolFont{AMSa}{U}{msa}{m}{n}
      \DeclareMathSymbol{\upi}{0}{UPM}{"19}
      \DeclareMathSymbol{\umu}{0}{UPM}{"16}
      \DeclareMathSymbol{\upartial}{0}{UPM}{"40}
      \DeclareMathSymbol{\leqslant}{3}{AMSa}{"36}
     \DeclareMathSymbol{\geqslant}{3}{AMSa}{"3E}
    \fi
  \fi
\fi 

\ifCUPmtlplainloaded \else
  \ifAMStwofonts \else 
    \def\upi{\pi}
    \def\umu{\mu}
    \def\upartial{\partial}
  \fi
\fi

\title[N/O impact on emission line diagnostics]{The impact of the nitrogen-to-oxygen ratio on ionized nebulae diagnostics based on [N{\sc ii}] emission lines.}
\author[E. P{\'e}rez-Montero \& T. Contini]
       {Enrique P{\'e}rez-Montero$^{1,2}$ \& Thierry Contini$^{2}$  \\
$^{1}$ Instituto de Astrof\'\i sica de Andaluc\'\i a, CSIC, Apdo. 3004, 18080, Granada, Spain.\\
$^{2}$ Laboratoire d'Astrophysique de Toulouse-Tarbes, Universit\'e de Toulouse, CNRS, 14, avenue Edouard Belin, F31400 Toulouse, France\\  }
        
\date{Accepted 
      Received ;
      in original form November 2006}

\pagerange{\pageref{firstpage}--\pageref{lastpage}}
\pubyear{2006}

\begin{document}

\maketitle

\label{firstpage}

\begin{abstract}
We study the relation between nitrogen and oxygen abundances
as a function of metallicity for a sample of emission-line objects for which
a direct measurement of the metallicity has been possible. This sample is representative of the very different conditions in
ionization and chemical enrichement that we can find in the Universe. We first construct 
the N/O vs. O/H diagram and we discuss its large dispersion at all metallicity regimes.
Using the same sample and a large grid of photoionization models covering very different values of the
N/O ratio, we then study the most widely used strong-line calibrators of metallicity based on [N{\sc ii}] emission lines, 
such as N2 and O3N2.
We demonstrate that these parameters underestimate the metallicity at low N/O ratios and viceversa. We investigate also 
the effect of the N/O ratio on different diagnostic diagrams used to discriminate narrow-line AGNs from star forming regions, 
such as the [O{\sc iii}]/H$\beta$ vs. [N{\sc ii}]/H$\alpha$, and we show that a large fraction of the galaxies catalogued as
composite in this diagram can be, in fact, star forming galaxies with a high value of the N/O ratio.
Finally, using strong-line methods sensitive to the N/O abundance ratio, like N2O2 and N2S2, 
we investigate the relation between 
this ratio and the stellar mass for the galaxies of the SDSS. We find, as in the case of the mass-metallicity relation,
a correlation between these two quantities and a flattening of the relation for the most massive galaxies, 
which could be a consequence of the
enhancement of the dispersion of N/O in the high metallicity regime.

\end{abstract}


\section{Introduction}

The detection and measurement of emission lines in optical spectra are powerful tools for the derivation of 
the physical properties and the chemical status of star forming galaxies. 
In the star forming regions, the ionized gas absorbs the ultraviolet light coming from massive young stars and
reemits it under the form of bright emission lines in the optical. 
However, the number of the detected emission lines and the quality of their measurement affect
the methodology and the accuracy of the analysis of the properties of the host star forming regions. 
This is particularly true for the derivation of the metallicity.
For instance, in the Giant Extragalactic HII Regions and in the Blue Compact Dwarf Galaxies,
this analysis is based on the detection of the faint auroral lines, the so-called direct method, which allows to derive
electron temperatures of the ionized gas and 
the chemical abundances using
the intensities of the bright collisional forbidden emission lines. 

This procedure is much more accurate when the 
determination of the thermal struture of the gas is based on the measurement  of different electron temperatures, each one associated 
to different ionization regions ({\em e.g.} Garnett, 1992).

However, for the faint and distant galaxies and also for the metal-rich ones,
the metallicity can only be derived using calibrators based on bright collisional emission lines. 
Many studies focused
on the analysis of different calibrators of nebular metallicity based on strong
lines ({\em e.g.} P\'erez-Montero \& D\'\i az, 2005; Kewley \& Ellison, 2008).
The calibration of these parameters can be empirical ({\em i.e.} using
direct measurements of the oxygen abundance in the Local Universe) or theoretical
({\em i.e.} using photoionization models covering different physical properties).
Unfortunately, the dispersion and the range of valid metallicity of each parameter 
depends strongly on the sample or on the input conditions of the models used to
calibrate them.

Another issue, not well explored so far, appears with the strong-line methods aiming 
to derive the oxygen abundance with emission lines of other elements (nitrogen, sulphur, etc). 
These calibrations can indeed depend on the relative abundance ratio of these elements.
This is the case, for instance, for the S$_{23}$ parameter (D\'\i az \& P\'erez-Montero, 2000;
based on sulphur emission lines) which has been defined to compute oxygen abundances. 
This is also the case for some other calibrators based on [N{\sc ii}] emission lines, like the N2 parameter
(among others Storchi-Bergmann et al., 1994; Van Zee et al., 1998 or Denicol\'o et al., 2002, hereafter DTT02) or the O3N2 parameter 
(Alloin et al., 1979; Pettini \& Pagel, 2004, hereafter, PP04), which are used to derive the total abundance of oxygen.

There are some disagreements about the constancy of the ratio of different elements, even in the case when both 
have a primary origin, like oxygen and sulphur ({\em e.g.} Garnett, 2002.). 
In the case of the N/O ratio, which involves a primary element, oxygen, and another one with an
extra secondary origin, like nitrogen, things are even more complicated. It is predicted
that in the low metallicity regime 
most of the nitrogen has a primary origin, coming mainly from massive stars,
while at higher metallicities the secondary production of nitrogen, coming mainly from low and intermediate mass stars, 
is dominant. 
Since the secondary production of nitrogen depends on the previous amount of oxygen stored in stars, via the CNO cycle, 
this implies a constant N/O ratio as a function of O/H
at low metallicities (Edmunds \& Pagel, 1978; Alloin et al., 1979) and a strong slope for higher metallicities.

However, numerous uncertainties remain regarding this scenario.
First, the relative amount of primary nitrogen produced in massive stars (Chiappini et al., 2005) and in low- and 
intermediate-mass stars is not well known.
This leads to an enhancement of the dispersion in both the low and the high metallicity regime. 
Second, the production of nitrogen depends sensibly on the stellar lifetimes and star formation efficiencies
as derived from chemical evolution models, which 
increases also the dispersion in the predictions of these models (Moll\'a et al., 2006).

Our main aim with this work is to explore the observational dispersion of the N/O vs. O/H relation using a wide sample of
HII regions and star-forming galaxies for which we derive consistently the oxygen and the nitrogen abundances
using the direct method.
We investigate also the effect of taking into account the real dispersion of this plot in the
input conditions of photoionization models. We study the impact of the variation of the N/O ratio on different strong-line calibrators of nebular 
metallicity and on the diagnostic diagrams used to
identify narrow line active galactic nuclei which are based on nitrogen emission lines. Finally, we apply our new calibrations
based on N/O to study the behaviour of the N/O ratio as a function of the stellar mass in the galaxies of the local Universe.
In the last section, we summarise our results.


\begin{table}
\begin{minipage}{85mm}
\vspace{-0.3cm}
\normalsize
\caption{Bibliographic references for the emission line fluxes of the compiled sample}
\begin{center}
\begin{tabular}{lcc}
\hline
\hline
Reference & Object type\footnote{Name of the object or the sample or type
of object in the sample: GEHR denotes Giant Extragalactic HII Regions;
HIIG, HII Galaxies and DHR, Diffuse HII Regions.} & Number \\
\hline
Bresolin et al., 2004 & M51 GEHRs & 10 \\
Bresolin et al., 2005 & GEHRs & 31 \\
Bresolin, 2007 & M101 GEHRs & 3 \\
Garnett et al., 2004 & M51 GEHRs & 2 \\
Guseva et al., 2003a & SBS 1129+576 & 2 \\
Guseva et al., 2003b & HS 1442+650 & 2 \\
Guseva et al., 2003c & SBS 1415+437 & 2 \\
H\"agele et al., 2006 & HIIG & 3 \\
H\"agele et al., 2008 & HIIG & 7 \\
Izotov \& Thuan, 1998 & IZw18  & 2 \\
Izotov et al. 1997 & SBS 0335-018 & 1 \\
Izotov et al., 1999 & IZw18 & 1 \\
Izotov et al., 2001 & Tol 65 & 1 \\
Izotov et al., 2004 & HIIG & 3 \\
Izotov \& Thuan, 2004 & HIIG & 33 \\
Kehrig et al. 2004 & HIIG & 24 \\
Kniazev et al. & SDSS galaxies & 12 \\
Lee et al., 2004 & KISS galaxies & 13 \\
Melbourne et al., 2004 & KISS galaxies & 12 \\
P\'erez-Montero \& D\'\i az, 2005 & All & 300 \\
Van Zee, 2000 & UGCA92 & 2 \\
Vermeij et al., 2002 & DRH & 9 \\
\hline
Total &  & 475 \\

\hline
\hline
\end{tabular}
\end{center}
\label{refs}
\end{minipage}
\end{table}


\section{Description of the sample}

We compiled from the literature a sample of emission-line objects with available 
measurements of [O{\sc ii}],
[O{\sc iii}] and [N{\sc ii}] bright emission lines and, at least, one auroral 
emission line with high signal-to-noise ratio in order to estimate electron temperatures and to derive ionic abundances
using the direct method.

This compilation includes HII regions in our Galaxy
and the Magellanic Clouds, Giant Extragalactic HII regions (GEHR), and HII Galaxies. 
Most of these objects belong to the compilation by P\'erez-Montero \& D\'\i az (2005) 
which was complemented with the sources listed in Table \ref{refs}.
The compilation contains 271 HII galaxies, 161 GEHRs and 43 HII regions
of the Galaxy and the Magellanic Clouds, what gives a total of 475 objects.


\section{Photoionization models}

A large grid of photoionization models was calculated to investigate the impact of
the variation of the N/O abundance ratio on the estimation of several physical properties of 
HII regions, like electron temperatures, ionization correction factor (ICF) or empirical parameters
based on [N{\sc ii}] emission lines.

We used the photoionization code Cloudy 06.02 (Ferland et al., 1998), taking as ionizing
source the spectral energy distributions (SEDs) of O and B stars with the code WM-Basic (version 2.11
\footnote{Available at http://www.usm.uni-muenchen.de/people/adi/Programs/Programs.html}, Pauldrach et al., 2001).
The use of the SEDs of single stars allows to quantify the effective temperature of the models and does not affect 
strongly to the ionization structure of the gas as compared with SEDs from massive clusters.
We assumed in all the models a radiation-bounded spherical geometry, a constant density of 100 particles per cm$^{2}$ and a
standard fraction of dust grains in the interstellar medium. 
In all the models the calculation was stopped when the temperature was lower than 4000 K.
We also assumed that the gas had the same metallicity than the ionizing stars,
covering the values 0.05\,Z$_\odot$, 0.2 Z\,$_\odot$, 0.4\,Z$_\odot$, Z$_\odot$ and 2\,Z$_\odot$, 
considering as the solar oxygen abundance the value measured by Allende-Prieto et al. (2001; 12+log(O/H) = 8.69). 
The other ionic abundances were set in solar proportions, except in the case of nitrogen, for which we considered values
for all metallicities relative to oxygen of -0.50, -0.75, -1.0, -1.25, -1.50 and -1.75 in
logarithmic units. Some depletion was taken into account for the chemical elements 
(C, O, Mg, Si and Fe) involved in the formation of dust grains. Regarding the other functional 
parameters we considered different values for the ionization parameter (log U = -3.5, -3.0, 
-2.5 and -2.0) and the effective temperature (T$_*$ = 40000 K, 45000 K and 50000 K). 
This produced a total of 360 photoionization models to cover the physical conditions of different 
ionized gas nebulae.  Atomic data, including collision strengths 
in the models were consistent with those used in the calculation of chemical abundances in the
compiled sample. They can be found in Table 6 of H\"agele et al. (2008).
We checked the influence of varying dielectronic recombination rate coefficients according with the most
recent values (Badnell, 2006 and references therein), but we did not find any significant variation in
our results.

\section{Results and discussion}

\subsection{Physical conditions: electron density and temperature}

All physical conditions were derived using the appropriate ratios of emission lines with
the tasks TEMDEN, in the case of electron densities and temperatures, 
and IONIC, in the case of ionic abundances. These IRAF \footnote{IRAF, Image Reduction and 
Analysis Facility, is distributed by the National Optical Astronomical Observatory} tasks are based on the five-level statistical
equilibrium model (De Robertis, Dufour \& Hunt, 1987; Shaw \& Dufour, 1995).

Following the procedure described in P\'erez-Montero \& D\'\i az (2003), we assigned 
the value of each derived electronic temperature to the calculation of the abundance of the 
corresponding ion. 
Electron densities were calculated for a subsample of 454 objects using the ratio of 
[S{\sc ii}] emission lines I(6717\AA)/I(6731\AA).
We assumed a density of 100 particles per cm$^2$, typical in this kind of objects, in those
objects without this ratio.

The electron temperature of O$^{2+}$ was derived using the ratio of [O{\sc iii}] emission lines 
(I(4959\AA)+I(5007\AA))/I(4363\AA) for a subsample of 418 objects.
For the objects that did not present the [O{\sc iii}] auroral line, we derived t([O{\sc iii}]) from 
the available line temperatures of other ions: 33 from t([S{\sc iii}]), using the empirical expression
by H\"agele et al. (2006), 20 from t([O{\sc ii}]), and 4 from t([N{\sc ii}])

The electron temperature of O$^+$ was derived 
using the [O{\sc ii}] emission line ratio (I(3727\AA)/(I(7319\AA)+I(7330\AA)) for a subsample of 196 objects.
The measurement of [O{\sc ii}] auroral lines presents higher uncertainties than in the case of [O{\sc iii}] 
due to their lower signal-to-noise ratio and 
to a higher dependence on reddening correction, caused by their larger wavelength distance to the closest hydrogen 
recombination line. 
Besides, the [O{\sc ii}] auroral lines are contaminated by a small contribution of dielectronic recombination emission,
although it does not contribute in quantities larger than the usual reported errors. 
Additionally, this ratio depends on electron density. 

All these factors can enhance the uncertainty associated to the total chemical abundance of oxygen, especially
in high metallicity - low excitation HII regions (P\'erez-Montero \& D\'\i az, 2005), but the
estimation of O$^+$ using its corresponding measured electron temperature allows a better
quantification of the associated errors in these objects.

For the objects with no direct estimation of t[O{\sc ii}], we used the 
relation between t([O{\sc ii}]) and t([O{\sc iii}]) proposed by P\'erez-Montero \& D\'\i az (2003).
This relation based on models has the following expression:

\begin{equation}
t([O{\sc ii}]) = \frac{1.2+0.002 \cdot n + \frac{4.2}{n}}{t([O{\sc iii}])^{-1} + 0.08 + 0.003 \cdot n + \frac{2.5}{n}}
\end{equation}

\noindent and they take into account the dependence of t([O{\sc ii}]) on the electron density, $n$.


\begin{figure}
\begin{minipage}{85mm}
\psfig{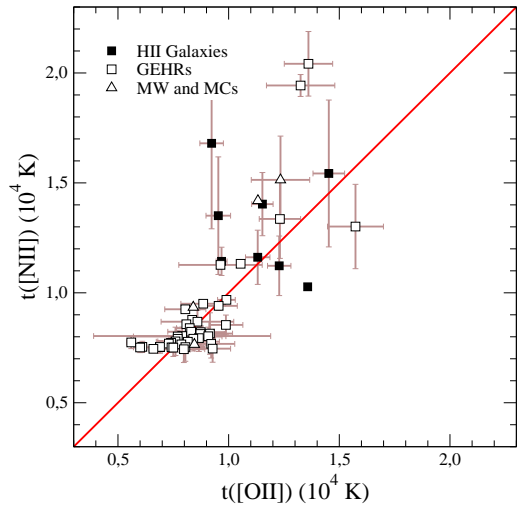}
\caption{Relation between t([O{\sc ii}]) and t([N{\sc ii}]) for the sample of objects with a direct measurement of both temperatures.
Black circles represent HII galaxies, white squares: Giant Extragalactic HII Regions, white triangles: HII regions in the Galaxy and the Magellanic
Clouds. The solid red line represents the 1:1 relation.}
\label{temp_01}
\end{minipage}
\end{figure} 


\begin{figure}
\begin{minipage}{85mm}
\psfig{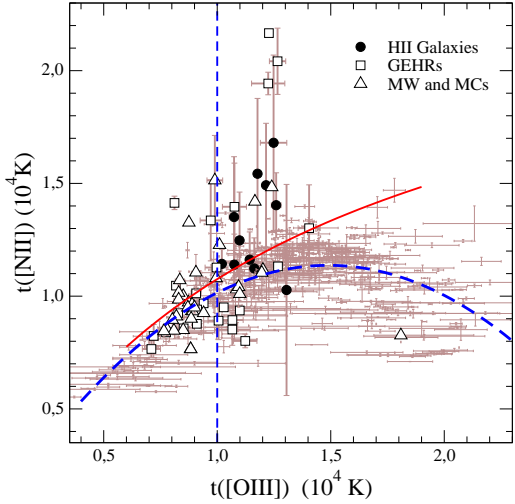}
\caption{Relation between t([O{\sc iii}]) and t([N{\sc ii}]) for the sample of objects with a direct measurement of both temperatures.
Symbols are the same as in the previous figure. 
The solid red line represents
the best quadratic fit to the photoionization models. The brown error bars represent the values of t([N{\sc ii}]) derived from the
relation of Thurston et al. (1996) and the dashed blue line its best quadratic fit. Vertical dashed line shows the upper limit (10000 K) 
of validity originally defined for the Thurston et al. relation.}
\label{temp_02}
\end{minipage}
\end{figure} 

Regarding nitrogen, the electron temperature of N$^+$ can be derived from the emission-line ratio (I(6584\AA)+I(6548\AA))/I(5755\AA)
in very few objects, due to the intrinsic weakness of the corresponding auroral line. We derived this temperature
for a subsample of 88 objects. 
In Figure \ref{temp_01}, we show the relation between t([O{\sc ii}]) and t([N{\sc ii}]) for those objects of the sample with
a direct determination of both temperatures. As we see, the assumption of a unique low-excitation electron temperature is
well justified for the most part of the objects. Nevertheless, the dispersion in this diagram and the
deviation between these temperatures in objects with thermal structures affected by the geometry of the gas lead
us to take independent estimations of these temperatures in all cases.

Therefore, for those objects without a direct estimation of t([N{\sc ii}]), we derived an independent 
relation between this temperature and t([O{\sc iii}]) 
based on the grid of photoionization models described in Section 3.
The obtained relation is well described by the following quadratic fit:
\begin{equation}
t([N{\sc ii}]) = \frac{1.85}{t([O{\sc iii}])^{-1}+0.72}
\end{equation}

This relation does not depend on the assumed N/O ratio in the range of temperatures covered
by the models (from 6000 to 18000 K).
We show it in Figure \ref{temp_02} together with the objects with a direct 
determination of both t([O{\sc iii}])  and t([N{\sc ii}]). As we can see, the relation deduced from the
models overestimates t([N{\sc ii}]) at very low temperatures as compared with the direct measurements.
However, in the range of T([O{\sc iii}]) between 7000 and 14000 K,
the relation covers correctly the sample, with the exception of some objects with
very high values of t([N{\sc ii}]).
This could be due to different problems in the measurement of the weak auroral line of [N{\sc ii}].
In these cases, we took the temperature predicted by the models to calculate nitrogen abundances.

We show in the same Figure \ref{temp_02} the temperatures of [N{\sc ii}] based on the relation proposed by 
Thurston et al. (1996), based also on models, for the high metallicity range (Z$>0.3$Z$_\odot$)

\begin{equation}
t([N{\sc ii}]) = 0.6065 + 0.1600x+0.1878x^2+0.2803x^3
 \end{equation}

\noindent where $x$ is based on the intensities of the emission lines of [O{\sc ii}] and [O{\sc iii}]

\begin{equation}
x = \log R_{23} = \log \left( \frac {I([OII] 3727 + [OIII] 4959,5007)}{I(H\beta)} \right)
\end{equation}

The values of t([N{\sc ii}]) deduced from this relation are plotted in Figure \ref{temp_02} with the best
quadratical fit.
This relation is almost identical to ours for low excitation conditions 
(T$<$10000 K), which is the range of validity 
originally proposed. 
The two relations diverge for higher temperatures, leading to 
non-negligible deviations in the calculation of the ionic abundances if it is not used properly.

\subsection {Ionic and total chemical abundances}

The ionic abundances were calculated using the most prominent emission lines of each ion and the appropriate
electron temperature. To calculate O$^+$ ionic abundances, we used the intensity of the [O{\sc ii}] line
at 3727 {\AA}, except in the case of 11 of the 12 very low metallicity HII Galaxies identified by Kniazev et al. (2003) 
in the SDSS catalogue 
without any measurement of the [O{\sc ii}]$\lambda$3727 line.
For these objects the O$^+$ abundance was derived using the emission lines of [O{\sc ii}]
 at 7319 \AA\ and 7330 \AA.
The derivation of O$^{2+}$ abundances was based on the [O{\sc iii}] emission lines at 4959 {\AA} and 5007 {\AA}.
The O$^+$ and O$^{2+}$  abundances were then used to calculate the total abundance of oxygen relative to hydrogen 
using the following assumption:

\begin{equation}
\frac{O}{H} \simeq \frac{O^++O^{2+}}{H^+}
\end{equation}


\begin{figure*}
\begin{minipage}{170mm}
\centerline{
\psfig{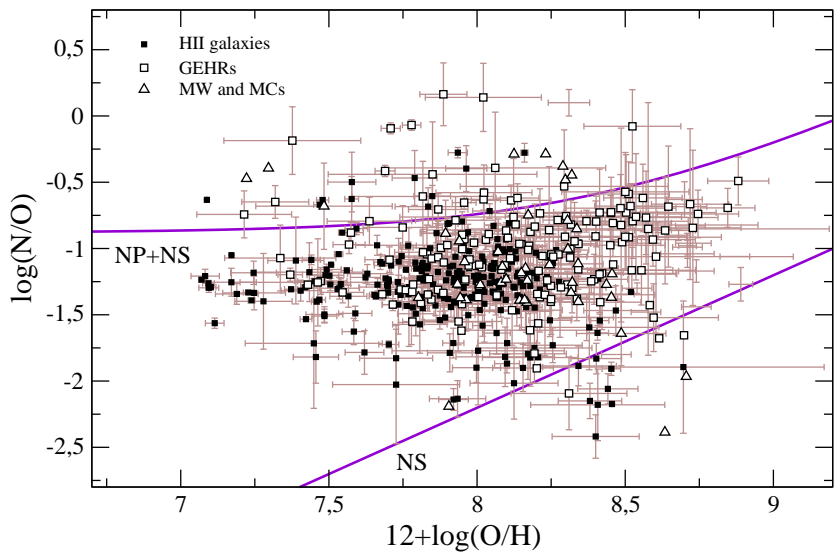}}
\caption{Relation between metallicity, represented by the abundance of oxygen, and the N/O ratio for different objects: the sample of
HII regions whose abundances were calculated using the direct method in this paper are represented by the same symbols
as in Figure 1. Solid line represents the prediction of chemical evolution closed-box models of primary and secondary nitrogen 
at all ranges of metallicity. These models were calculated taking the yields from Henry et al. (2000) }
\label{no_o}
\end{minipage}
\end{figure*} 

Abundances of N$^{+}$ were calculated using the [N{\sc ii}] emission lines at 6548 {\AA} and 6584 {\AA}
Then it is possible to derive the total nitrogen abundance and the N/O ratio taking the following approximation:

\begin{equation}
\frac{N}{O} \simeq \frac{N^+}{O^+}
\end{equation}

\noindent based on the similarity of the ionization structures of both N and O.  This leads to
the following assumption about the ICF of N$^+$:

\begin{equation}
ICF(N^+) \simeq \frac{O}{O^+} \simeq \frac{O^++O^{2+}}{O^+}
\end{equation}

\noindent which is in good agreement with our models.

\subsection{Behaviour of N/O with metallicity}

We show in Figure \ref{no_o} the relation between the N/O ratio and the 
oxygen abundance for the sample of
HII regions {and star-forming galaxies
whose oxygen and nitrogen abundances were calculated consistently following the direct method (see Section 2).

We also show the prediction from closed-box chemical evolution models concerning the production of 
secondary nitrogen (yields from Henry et al., 2000), in the high metallicity regime, and the contribution of an extra amount of
primary nitrogen at lower metallicities.

As we can see, although these predictions are followed in average, the observations show a 
very high dispersion, especially at high metallicities. 
This dispersion is real and does not depend on the methodology we have used to derive abundances.
For instance, the assumption of a single electron temperature for the calculation of N$^+$ and O$^+$
only reduces the standard deviation in the distribution of N/O in 0.05 dex.
At low metallicities,
the dispersion is lower because nitrogen has essentially a primary origin, produced mainly
by massive stars and with no dependence on the previous amount of oxygen stored in stars.
However, even in this regime, there are some objects which exceed the relative amount
of nitrogen predicted by the models. This is the case of some HII galaxies and GEHRs, with a
probable extra production of primary nitrogen. 
Between the causes of the uncertainty in the prediction of primary N, it is
the rotation velocity of stars and the amount of metals loss in winds, as in Wolf-Rayet stars (Meynet \& Maeder, 2005).

For high metallicities the dispersion increases as a consequence of both primary
and secondary production of nitrogen in the same metallicity regime. It is known that some 
intermediate-mass stars can produce an extra amount of primary nitrogen ({\em e.g.} Renzini \& Voli, 1981). 
We also find some HII galaxies with extremely low values of N/O. This type of objects was already described in
some works in which they are often associated with UV-bright galaxies ({\em e.g.} Contini et al., 2002;
Mallery et al., 2007).
For the same metallicity, some HII regions in spirals show very high values of the N/O ratio
but, in this case, Bresolin et al. (2005) claim that the N abundances determined
from the [N{\sc ii}] auroral line are not very reliable.

The simultaneous analysis of all these objects in the same diagram leads to the 
conclusion that in the same metallicity bin, we can find variations in the N/O ratio of more than one 
order of magnitude and implies the impossibility of finding an unique model that predicts the behaviour of 
this ratio for all the metallicity regimes.


\begin{figure}
\begin{minipage}{85mm}
\psfig{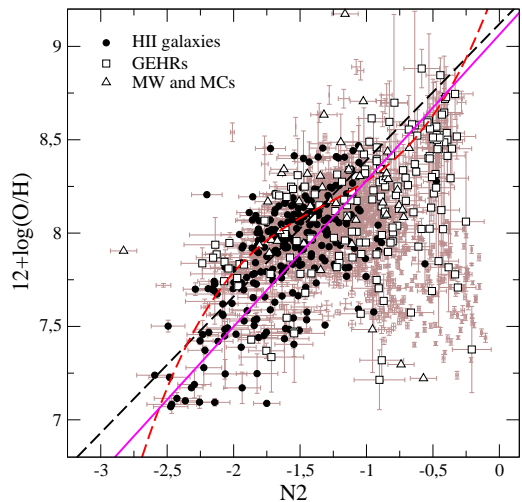}
\caption{Relation between the N2 parameter and the oxygen abundance for the studied objects.
Symbols stand for the same types of objects as described in previous figures. Black dashed line represents
the DTT02 calibration, red dashed line the PP04 calibration and magenta solid line the
linear fit to the data.}
\label{n2_o}
\end{minipage}
\end{figure} 


\begin{figure*}
\begin{minipage}{170mm}
\centerline{
\psfig{figure=N-to-O-pmc.fg05.eps,width=13cm,clip=}}
\caption{Relation between oxygen abundance and the N2 parameter for different values of the N/O ratio. All the sample is shown in
each plot but only those objects with the corresponding N/O value are marked as black squares.
We show too as white circles the photionization models corresponding to these 
values of N/O: -1.75, -1.50, -1.25, -1.00, -0.75 and -0.50 from right to left and from bottom to top, respectively.}
\label{n2_o_mod}
\end{minipage}
\end{figure*} 


\begin{figure}
\begin{minipage}{85mm}
\psfig{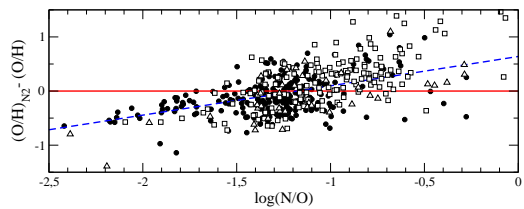}
\caption{Residuals between the oxygen abundances derived from the N2 empirical calibration and the abundances derived using the direct
method as a function of the N/O ratio. The solid line represents the residual zero and the dashed line the linear fit of the points.}
\label{n2_res}
\end{minipage}
\end{figure} 

\subsection{Metallicity strong-line methods based on [N{\sc ii}]}

In this section, we investigate the effect of the N/O abundance ratio on two metallicity 
empirical calibrators based on [N{\sc ii}] emission lines, namely N2 and O3N2.

\subsubsection{The N2 parameter}

The N2 parameter, defined as:

\begin{equation}
N2 = log \left( \frac {I([N{\sc ii}] 6584 \AA)}{I(H\alpha)}\right)
\end{equation}

\noindent has been extensively used in the literature to derive oxygen abundances.
({\em e.g.} Storchi-Bergmann et al., 1994, Van Zee et al., 1998, DTT02).
In particular, DTT02 proposed the following linear relation:

\begin{equation}
12+log(O/H) = 9.12 + 0.73 \cdot N2,
\end{equation}

\noindent based on a sample of HII regions with metallicities derived in the low metallicity regime using
the direct method and using other strong-line methods in the high metallicity regime. 
Another extensively used calibration of this parameter is this polynomical fit proposed by PP04.

\begin{equation}
12+log(O/H) = 9.37+2.03 \cdot N2+1.26\cdot N2^2 + 0.32\cdot N2^3
\end{equation}

The used sample in this case is composed of HII regions whose oxygen abundance were mostly derived using the direct method,
except in the case of some oversolar metallicity objects, whose metallicities were derived from photoionization models.

The main advantage of this parameter lies in its complete independence on reddening correction
or flux calibration, due to the close wavelength of the involved emission lines and, contrary to the
R$_{23}$ parameter, in its single-valued relation with metallicity up to solar values.
In Figure \ref{n2_o}, we show the relation between the N2 parameter and the oxygen abundance
for our sample of objects, along with the calibrations of DTT02
and PP04. The least-squares bisector linear fit (Isobe et al., 1990) of the points
is plotted in Figure \ref{n2_o} and it gives the following expression:

\begin{equation}
12+log(O/H) = 0.79 \cdot N2 + 9.07
\end{equation}

\noindent 
However, this fit does not provide a lower dispersion as compared to the previous calibrations: 
the standard deviations of the residuals is 0.34 dex, while it is 0.33 in the case of the DTT02 calibration and 0.32 dex
for PP04. The origins of this dispersion are studied in P\'erez-Montero \& D\'\i az (2005), who
concluded that 
higher ionization parameters and effective temperatures lead to higher metallicities as derived from N2.
Additionally, a substantial part of the observed dispersion is due to N/O variations, 
as a consequence of using nitrogen emission lines to derive oxygen abundances.

In Figure \ref{n2_o_mod}, we show the studied sample divided in bins of log(N/O),
and compared with sequences of photoionization models with the same values of N/O. 
These sequences of models have as input conditions the values of T$_*$ and log U most plausible
for each metallicity (high T$_*$ and log U for low metallicity and viceversa),
but they have not been used to recalibrate the parameter and they are only used
to check how the N2 parameter varies as a function of N/O in the models.


\begin{figure}
\begin{minipage}{85mm}
\centerline{
\psfig{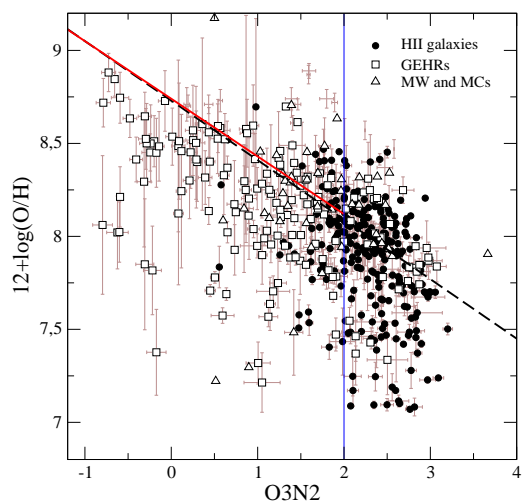}}
\caption{Relation between the O3N2 parameter and the oxygen abundance for the studied objects.
Symbols stand for the same types of objects as described in previous figures. Black dashed line represents
the PP04 calibration and blue vertical solid line shows the limit of validity of this calibration (O3N2 $<$ 2)}
\label{o3n2_o}
\end{minipage}
\end{figure} 

\begin{figure*}
\begin{minipage}{170mm}
\centerline{
\psfig{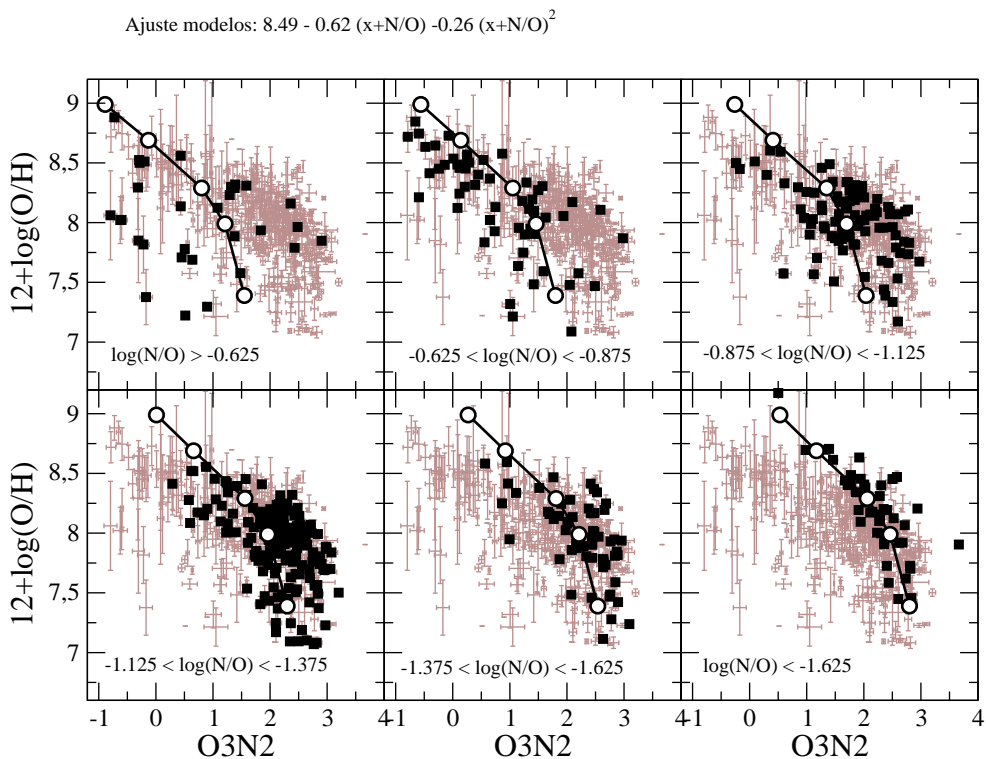}}
\caption{Relation between oxygen abundance and the O3N2 parameter for different values of the N/O ratio. 
The full sample is shown in
each plot but only those objects with the corresponding N/O value are marked as black squares.
We show also as white circles the photionization models corresponding to these 
values of N/O: -1.75, -1.50, -1.25, -1.00, -0.75 and -0.50 from right to left and from bottom to top.}
\label{o3n2_o_mod}
\end{minipage}
\end{figure*} 

As we can see, it
exists a strong correlation between the metallicity derived from the N2 parameter and the N/O ratio in such a way
that the metallicity predicted by N2 is overestimated in those objects with a high N/O ratio and vice-versa.
This trend is confirmed when we compare models 
and the observations, although the dispersion in the observations is higher for high values of N/O.
Regarding models, no consistent values were found for oversolar abundances because
the N2 parameter saturates in this high metallicity regime.

We quantified the dependence of the calibration of the N2 parameter on the N/O ratio.
To do so, we analysed the residuals between the metallicities derived from N2 and 
from the direct method as a function of N/O.
This is shown in Figure \ref{n2_res} with its best linear fit, which gives the following result:

\begin{equation}
\Delta(O/H) = 0.56 \cdot \log(N/O) + 0.66
\end{equation}

Taking into account this correction, we can adopt the following expression for the relation between 
N2 and the oxygen abundance:

\begin{equation}
12+\log(O/H) = 0.79 \cdot N2 - 0.56 \cdot \log (N/O) + 8.41
\end{equation}

\noindent which reduces the dispersion of the metallicity derived using N2 to 0.21 dex.

\subsubsection{The O3N2 parameter}

The O3N2 parameter is defined as:

\begin{equation}
O3N2 = \log \left( \frac{I([O{\sc iii}] 5007 {\AA)}}{I(H\beta)} \times \frac{I(H\alpha)}{I([N{\sc ii}] 6584 {\AA})} \right)
\end{equation}

\noindent and was used as an estimator of metallicity by Alloin et al. (1979), although
the most widely used calibration is a linear fit proposed by PP04:

\begin{equation}
12+log(O/H) = 8.73 - 0.32 \cdot O3N2
\end{equation}

We show in Figure \ref{o3n2_o} the relation between the O3N2 parameter and the oxygen abundances
for our sample. Contrary to N2, the relation is not clearly
linear. For the low metallicity regime (O3N2 $>$ 2) 
this parameter does not correlate with the metallicity.
This means that for HII galaxies, which populate
the low metallicity regime, objects with a difference of an order of magnitude in metallicity have very
similar values of O3N2.
However, in the high metallicity regime (O3N2 $<$ 2), we can perform a 
least-square bisector linear fit to our data, which gives:

\begin{equation}
12+log(O/H) = 8.74 - 0.31 \cdot O3N2
\end{equation}


\begin{figure}
\begin{minipage}{85mm}
\psfig{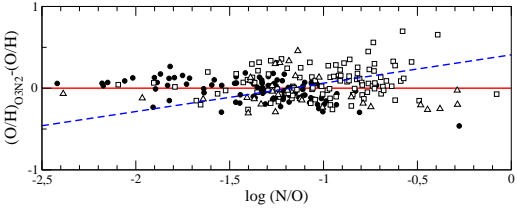}
\caption{Residuals between the oxygen abundances derived from the O3N2 empirical calibration and the abundances derived using the direct
method as a function of the N/O ratio. The solid line represents the residual zero and the dashed line the linear fit of the points.}
\label{o3n2_res}
\end{minipage}
\end{figure} 


\begin{figure*}[t]
\begin{minipage}{170mm}
\centerline{
\psfig{figure=N-to-O-pmc.fg10l.eps,width=6.8cm,clip=}
\psfig{figure=N-to-O-pmc.fg10r.eps,width=6.8cm,clip=}}
\caption{Relation between the N2O2 parameter and the oxygen abundance (left panel) or the N/O abundance ratio (right panel) for the objects described in Section 2. The solid line represents the best linear fit to the sample.
}
\label{n2o2}
\end{minipage}
\end{figure*} 


\begin{figure*}
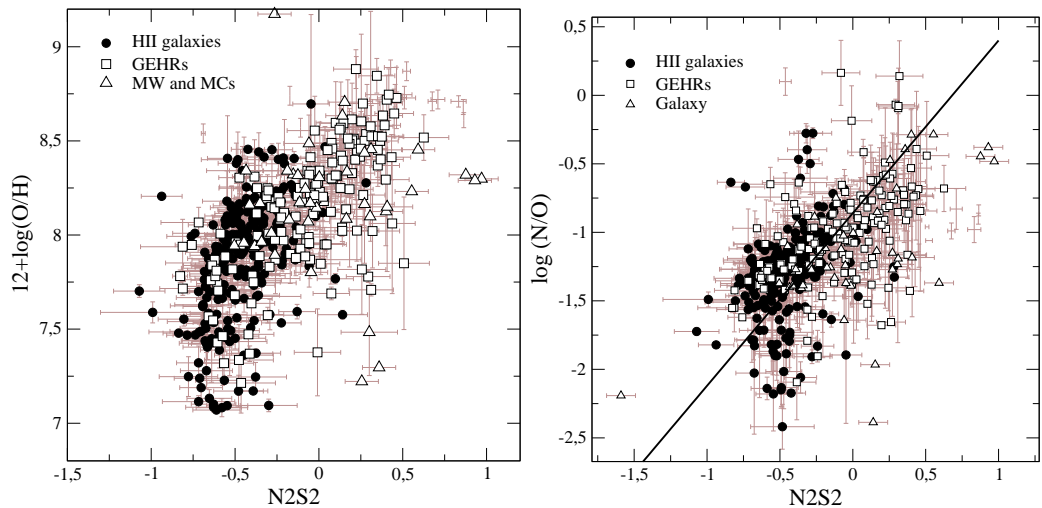

\begin{minipage}{170mm}
\centerline{
\psfig{figure=N-to-O-pmc.fg11l.eps,width=6.8cm,clip=}
\psfig{figure=N-to-O-pmc.fg11r.eps,width=6.8cm,clip=}}
\caption{Relation between the N2S2 parameter and the oxygen abundance (left panel) or the N/O abundance ratio (right panel) for the objects described in Section 2. The solid line represents the best linear fit to the sample.}
\label{n2s2}
\end{minipage}
\end{figure*} 

\noindent and which is almost identical to the relation proposed by PP04. Both relations
present a standard deviation of the residuals of 0.32 dex.

For the O3N2 parameter we show in Figure \ref{o3n2_o_mod} the same data but divided in bins corresponding to different 
N/O ratios. The models corresponding to different values of log(N/O) (-1.75, 
-1.50, -1.25, -1.00, -0.75 and -0.50) are shown too. 
Despite of the large dispersion found for high values of N/O, as in the case of N2, we
see a strong dependence of the relation on the N/O ratio for both the models and the observations.
This implies that the metallicity predicted by O3N2
is overestimated in those objects with a high value of the N/O ratio.

The linear relation existing between the O3N2 parameter and the oxygen abundance
is valid only for  12+log(O/H) $>$ 8 but, contrary to N2 parameter,
it can be used up to very high metallicities, larger than twice the solar value, according to the models.

We show in Figure \ref{o3n2_res}, the residuals between the metallicities derived from O3N2 and the
abundances obtained following the direct method in the high
metallicity regime (12+log(O/H) $>$ 8) as a function of the N/O ratio. 
The least-square bisector linear fit to the data points 
has the following expression:

\begin{equation}
\Delta (O/H) = 0.35 \cdot \log (N/O) + 0.41
\end{equation}

The use of this correction to the linear fit of O3N2 at high metallicities reduces the dispersion to 0.23 dex,
using the following expression:

\begin{equation}
12+\log(O/H) = 8.33 - 0.31 \cdot O3N2 -0.35 \cdot \log(N/O)
\end{equation}

\subsection{Strong-line methods to derive the N/O ratio}

The N2O2 parameter defined as:

\begin{equation}
N2O2 = \log \left( \frac{I([N{\sc ii}]6584{\AA})}{I([O{\sc ii}]3727{\AA})} \right)
\end{equation}

\noindent has been proposed as an estimator of oxygen abundance by Kewley \& Dopita (2002).
In the left panel of Figure \ref{n2o2} we show the relation between this parameter and the oxygen abundance
for our sample. As it was already stated by Kewley \& Dopita (2002),
this parameter is sensitive to the metallicity in the high metallicity regime only, while for
low metallicity objects, the parameter is almost constant. In fact, the relation between N2O2
and metallicity resembles the relation between N/O and O/H because it reproduces the
ratio between the abundances of oxygen and nitrogen: a constant value of
the nitrogen abundance in the low metallicity regime and a strong variation as a function
of metallicity for high metallicities.

In the right panel of Figure \ref{n2o2} we show the relation between N2O2 and
the N/O ratio. In this case, we find a linear correlation for all metallicity regimes, with the
exception of some HII galaxies with very low values of N/O.
The least-squares bisector linear fit between this parameter and the N/O ratio gives

\begin{equation}
\log(N/O) = 0.93 \cdot N2O2 - 0.20
\end{equation}

\noindent with a standard deviation of the residuals of 0.24 dex. 
This relation is slightly shallower than the linear fit proposed by
P\'erez-Montero \& D\'\i az (2005) for a sample of HII galaxies.
Therefore, the ratio between [N{\sc ii}] and
[O{\sc ii}] lines constitutes a powerful tool to derive N/O ratio. Moreover it allows 
the correction of the metallicity estimated from the N2 parameter or,
in the case of the oversolar metallicity regime, from the O3N2 parameter.

Another calibrator based on [N{\sc ii}] is the S2N2 parameter. Originally introduced by
Sabbadin et al. (1977) it has been proposed to derive metallicities by Viironen et al. (2007).
To keep the consistency with previous described parameters, we define
this calibrator as follows:

\begin{equation}
N2S2 = \log \left( \frac{I([N{\sc ii}] 6584{\AA})}{I([S{\sc ii}] 6717,6731{\AA})} \right)
\end{equation}

In the left panel of Figure \ref{n2s2}, we show the relation between this parameter and
the abundance of oxygen in our sample. We find again the same trend observed
in the case of N2O2: a constant value of the parameter for low metallicities and a variation 
for high metallicities. This is a consequence of using a ratio of emission lines coming
from sulphur, which is an $\alpha$ element whose origin is primary, as in the
case of oxygen,
and nitrogen, whose origin can be primary or secondary. 
As for N2O2 we explored the correlation between this
parameter and the N/O ratio, which is shown in right panel of Figure \ref{n2s2}. This
relation is again linear with the exception of some HII galaxies with very low values of N/O.
The least-squares bisector linear fit of this parameter 
as a function of the N/O ratio gives:
\begin{equation}
\log(N/O) = 1.26 \cdot N2S2 - 0.86
\end{equation}

\noindent with a standard deviation of the residuals of 0.31 dex, a slightly higher dispersion
than in the case of N2O2 caused probably by the variations in the S/O ratio, taking into
account that uncertainties 
due to reddening and flux calibrations are lower than in the case of N2O2.
Nevertheless, N2S2 is more suitable to estimate the N/O ratio than N2O2 due to
the simultaneous measurement of all the involved lines in many observations.


\begin{figure*}[h]
\begin{minipage}{170mm}
\centerline{
\psfig{figure=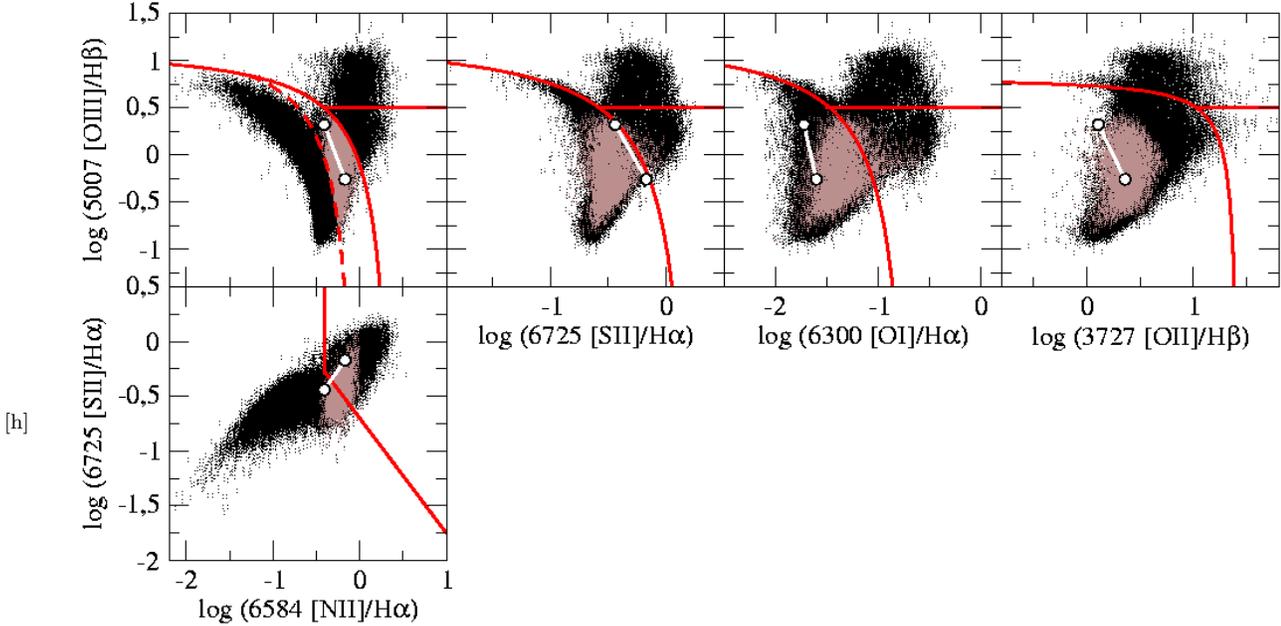,width=16cm,clip=}}

\caption{Different diagnostic diagrams used to separate star forming galaxies and narrow-line AGNs. We show in the four top panels
the relations between [O{\sc iii}]/H$\beta$ and from left to right: [N{\sc ii}]/H$\alpha$, [S{\sc ii}]/H$\alpha$, [O{\sc i}]/H$\alpha$
and [O{\sc ii}]/H$\beta$. The bottom panel represents the relation between [N{\sc ii}]/H$\alpha$ and [S{\sc ii}]/H$\alpha$. Brown points
represent {\em composite} galaxies (see text) whose log(N/O) as derived from the N2O2 parameter is higher than -0.5. 
Solid lines represent the curves proposed by Kewley et al. (2003) to separate
star forming galaxies  (bottom left) and AGNs (top right) stay. The dashed line in the first panel represents the empirical curve predicted by Kauffmann et al. (2003). Finally, the white points represent two different models of star
forming regions with high metallicity (Z=0.02Z$_\odot$) and high N/O ratio ($-0.5$).}
\label{agns}
\end{minipage}
\end{figure*} 

\subsection{Spectral classifications based on [N{\sc ii}] lines}

Diagnostics based on [N{\sc ii}] emission lines are commonly used to
classify emission-line galaxies according to their dominant
mechanism of ionization.
In particular, it is possible to distinguish between 
ionization due to the UV radiation coming from massive young stars (star-forming regions) 
or from an active galactic nucleus (hereafter AGN).
Among these diagnostics, the first diagrams proposed by Baldwin, Philips \& Terlevich 
(1981), also known as BPT diagrams, are based on four emission-line ratios, namely
[O{\sc iii}]/H$\beta$, [N{\sc ii}]/H$\alpha$, [S{\sc ii}]/H$\alpha$ and [O{\sc i}]/H$\alpha$.
Due to the impossibility of detecting at the same time these emission lines in most of the deep 
surveys of high-redshift galaxies based on optical spectroscopy (eg. VVDS, zCOSMOS), because 
of their limited spectral range, some other diagnostic diagrams were proposed to overcome this issue. 
This is the case of the blue diagnostic diagram, proposed by Lamareille et al. (2004), which involves 
the [O{\sc iii}] and [O{\sc ii}] lines} or the red diagnostic, which relates empirically the 
[N{\sc ii}]/H$\alpha$ and [S{\sc ii}]/H$\alpha$ ratios.

Generally in these diagnostic diagrams, the regions of star-forming galaxies and 
narrow-line AGNs (Seyfert 2 and LINERs) are separated using photoionization models 
(e.g. Kewley et al. 2001), or empirically (Kauffmann et al., 2003). 
In Figure \ref{agns} we show the diagnostic diagrams mentionned 
above using the emission-line galaxies listed in the MPA/JHU Data catalogue of the 
Sloan Digital Sky Survey\footnote{Available at http://www.mpa-garching.mpg.de/SDSS/} DR7 release. 
We have kept only the galaxies with a 
signal-to-noise ratio of at least 5 in all the 
involved lines. This excludes all the objects at redshift z$<$0.02, whose [O{\sc ii}] 3727 {\AA}
emission line is not observed in the SDSS catalogue and giving as a result
a total of 101753 emission line galaxies. 

We corrected for reddening 
the emission-line intensities using the Balmer decrement.
In the diagram [O{\sc iii}]/H$\beta$ vs. [N{\sc ii}]/H$\alpha$, the
separation curves proposed by Kewley et al. (2001) and Kauffmann et al. (2003) do
not coincide. In fact, as it is described in Kewley et al. (2006), the number of star-forming galaxies
predicted by the empirical separation is overestimated when compared with the
theoretical relation. This is explained in terms of a population of {\em composite} objects,
whose ionization is due partially to the star formation and to the presence of
an active nucleus. This agrees with the fact that [N{\sc ii}] emission lines are more
sensitive to the presence of X-ray sources and, therefore, they allow to probe the presence
of low-activity active nuclei in some galaxies classified as star-forming.

Nevertheless, the enhancement of the N/O abundance ratio due to peculiar chemical
histories can lead to high values of the [N{\sc ii}] emission lines in some star-forming galaxies. 
In all panels of Figure \ref{agns}
we identify those SDSS galaxies catalogued as {\em composite} following Kewley's criterion and
whose log(N/O) is higher than $-0.5$
as derived using the N2O2 parameter. This represents a 53\% of all
the {\em composite} galaxies.
At same time, most of these {\em composite}-high N/O galaxies lie in the bottom and left-hand side of
the Kewley et al. (2003) classification of the other diagrams:
86\% in the [S{\sc ii}]/H$\alpha$ diagram, 76\% in the [O{\sc i}]/H$\alpha$
and 99.8\% using [O{\sc i}]/H$\beta$.
This indicates that a fraction of the galaxies classified as {\em composite} in
the {N{\sc ii}]/H$\alpha$ vs. [O{\sc iii}]/H$\beta$ diagram could correspond
to star-forming galaxies with high values of the N/O ratio.

This conclusion is partially supported by some of the results coming from
photoionization models.
In Figure~\ref{agns} we show also the position of two of the models
described in Section 3. In this case, these models have an ionization coming from a
WM-Basic star with T$_*$=45000 K, an ionization parameter of respectively
log U = $-2.5$ and $-3.0$, a solar metallicity and a log(N/O) = $-0.5$. These models, which
appear in the {\em composite} region in the diagram {N{\sc ii}]/H$\alpha$ vs.[O{\sc iii}]/H$\beta$, are located below and in
the left-hand side of the Kewley et al. (2003) classification in the rest of the BPT diagrams, 
and in the blue diagnostic diagram too.
Only in the red diagnostic 
diagram between [N{\sc ii}]/H$\alpha$ and [S{\sc ii}]/H$\alpha$, the model
with a lower ionization parameter lies in the AGN region. 
We did not find
that the number of AGNs predicted by this red diagnostic diagram is higher than in the others not depending on [N{\sc ii}] lines,
but it is possible that the AGN region was contaminated with some star-forming objects with high N/O.


\begin{figure*}[h]
\begin{minipage}{170mm}
\centerline{\psfig{figure=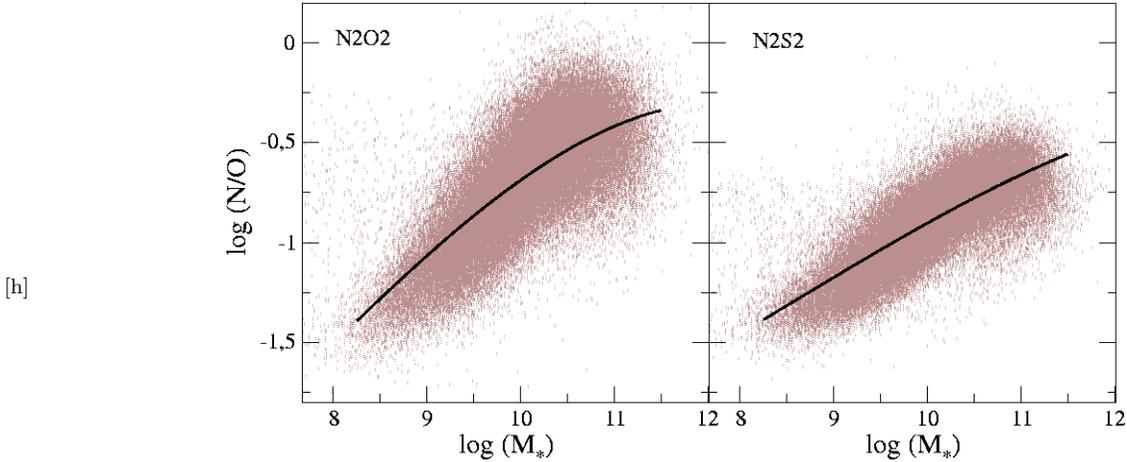,width=12cm,clip=}}
\caption{Relation between stellar mass and the N/O ratio for the star forming galaxies of the SDSS DR7 catalogue.  In the left
panel N/O ratios have been derived using the N2O2 parameter and in right panel, from N2S2.
The solid lines in both panels represent the polynomial fit of 3th order to the sample.}
\label{mNO}
\end{minipage}
\end{figure*} 

\subsection{The relation between the N/O abundance ratio and the stellar mass}

As an application of the strong-line methods based on [N{\sc ii}] emission lines to derive N/O, we studied the relation between this abundance ratio 
and the stellar mass of the star-forming galaxies selected in the DR7 of the SDSS. 
The stellar masses taken from the DR7 SDSS catalogue were derived from their SED fitting 
(Kauffmann et al., 2003). 
As before, we kept only those galaxies whose [O{\sc ii}], H$\alpha$, [N{\sc ii}] and [S{\sc ii}] emission-line fluxes 
had S/N better than 5.  This excludes as well the galaxies with redshift lower than 0.02, with no
measurement of the [O{\sc ii}] emission line.
We also excluded narrow-line AGNs as identified in the [O{\sc iii}]/H$\beta$ vs. [N{\sc ii}]/H$\alpha$ diagnostic diagram,
according to the Kauffman et al. (2003) curve and, therefore, we take all the {\em composite} galaxies.

The relation between N/O and stellar masses for SDSS galaxies is shown in Figure \ref{mNO}. 
In the left panel, the N/O ratios were derived using the
N2O2 parameter, while in the right panel they were obtained from N2S2.
The N/O values in this last case are,
in average, 0.25 dex lower. 
This difference is much higher
than the expected associated error to the lines, to the calibrations or to the reddening, so it is
possible that [O{\sc ii}] emission lines were underestimated in these objects due to their
proximity to the instrumental edge in SDSS observations.
However, we observe in both panels a global trend for a higher N/O ratio at
higher stellar masses.  This could be a consequence of the fact that the most massive galaxies
have evolved more quickly and, hence, they have in average higher metallicities and higher
N/O ratios.
We also observe a flattening of the relation
for the most massive galaxies. This is caused mainly by the same flattening
observed in the mass-metallicity relation (Tremonti et al., 2004), but it can be related to the
high dispersion in the N/O vs. O/H diagram  
at high metallicities, where the relative production of primary and secondary nitrogen is
quite uncertain. Although the variation in the N/O ratio can be higher than an order of
magnitude for each mass bin, we provide the cubic polymial fits to the data in
both diagrams to be used as a reference of the local Universe. 
In the case of the N/O ratio obtained using N2O2, we obtain the following relation: 

\begin{equation}
y = -1.884-0.846\cdot x + 0.172 \cdot x^2 - 0.0075 \cdot x^3
\end{equation}

and for N/O ratios obtained from the N2S2 parameter:

\begin{equation}
y = -1.5793-0.467\cdot x + 0.088 \cdot x^2 - 0.0034 \cdot x^3
\end{equation}

\noindent where $y$ is log(N/O) and $x$ is log(M$_*$), in solar units.

\section{Summary and conclusions}

In this work we investigated the impact of the N/O ratio on some strong-line methods commonly used to derive 
the metallicity of star-forming galaxies, and on different diagnostic diagrams based on [N{\sc ii}] 
emission lines.

To this aim, we compiled a sample of ionized gaseous nebulae with available measurements of 
the [O{\sc ii}], [O{\sc iii}] and [N{\sc ii}] strong lines and, at least, one auroral line. 
This allowed us to derive electron temperatures and ionic abundances following
the direct method.  The collected sample corresponds to very different metallicities and ionization conditions.
We also calculated a large grid of photoionization models covering these many
different conditions, including a range in log(N/O) between $-1.75$ and $-0.50$. As expected, we did not find any
significant dependence on N/O, both for the ICF(N$^+$) and for the relation between t([O{\sc iii}]) and t([N{\sc ii}]).
The inner thermal structure of the ionized gas is controlled mainly by the 
total metal content of the nebula and the relative abundances of different elements has a negligible impact on it.

The study of the relation between the N/O abundance ratio and the metallicity for the same sample 
confirmed a very high dispersion for
both the low and high metallicity regimes, due to the uncertainties in the relative production of primary and secondary
nitrogen at all metallicity regimes. We also studied the dependence of the widely used strong-line methods
N2 and O3N2 on the N/O ratio. Both observations and models confirm that these two parameters underestimate
the oxygen abundance for objects with low N/O ratios and viceversa.  Taking into account the
value of the N/O ratio in the calibrations can reduce 0.10 dex the dispersion of the residuals.
However, this dispersion cannot be reduced further due to the intrinsic heterogeneity in the considered sample.

The strong-line calibrators which depend on ratios between the [N{\sc ii}] emission line and
another low-excitation line, like [O{\sc ii}] or [S{\sc ii}]  (namely N2O2 and N2S2) can be used to derive 
the N/O ratio at all ranges of metallicity.

We explored the impact of the N/O ratio on the diagnostic diagram [O{\sc iii}]/H$\beta$ vs.
[N{\sc ii}]/H$\alpha$. We showed that the region of this diagram defined by Kewley et al. (2006)
for composite galaxies can also be populated by star-forming galaxies with a high N/O
ratio. More work in this issue must be done to
quantify the contribution of both populations to this part of the diagram.

Finally, by using the strong-line methods to derive N/O, we studied the variation
of this abundance ratio with the stellar mass of the SDSS galaxies. 
Despite the non-negligible offset between the ratios obtained using N2O2 and N2S2,
both diagrams are consistent when predicting
a enhancement of the average N/O for the most massive galaxies, consequence of
a more evolved status of these galaxies. This correlation is flatter for the most massive galaxies, 
which could be consequence of the same flattening observed in the mass-metallicity relation and of a
higher dispersion of the N/O ratio for higher metallicities.

\section*{Acknowledgements}
This work has been supported by the CNRS-INSU (France) and its Programme National Galaxies and the projects AYA-2007-767965-C03-02 and 03 of the Spanish National Plan for Astronomy and Astrophysics. Also, partial support from the Comunidad de Madrid under grant S0505/ESP/000237 (ASTROCAM) is acknoledged. EPM acknowledges his financial support from the {\em Fundaci\'on Espa\~nola para la Ciencia y la Tecnolog\'\i a} during its two-year postdoctoral position in the LATT-OMP.
We acknowledge Mercedes Moll\'a and Marta Gavil\'an for kindly providing us some of the data and the models to populate the N/O vs. O/H
and Fabrice Lamareille for his help for the analysis and treatment of SDSS galaxies. We would like to thank as well Jos\'e M. V\'\i lchez 
for very interesting discussions and comments that have help to improve this work and the anonymous referee
for his/her constructive comments.

\end{document}